\documentclass[10pt,A4paper]{article}

\usepackage{epsfig,amssymb}

\begin{document}

\title{\bf Fluctuation-Dissipation-Theorem violation during the
formation of a colloidal-glass}
\author{L. Bellon,  S. Ciliberto and  C. Laroche \\
         Ecole Normale
Sup\'erieure de Lyon, Laboratoire de Physique ,\\
 C.N.R.S. UMR5672,  \\ 46, All\'ee d'Italie, 69364 Lyon Cedex
07,  France\\
        }
 \maketitle

 \begin{abstract}
The relationship between the conductivity and the polarization
noise is measured in a gel as a function of frequency in the range
$1Hz - 40Hz$. It is found that at the beginning of the transition
from a fluid like sol to a solid like gel the fluctuation
dissipation theorem is strongly violated. The amplitude and the
persistence time of this violation are decreasing functions of
frequency. At the lowest frequencies of the measuring range it
persists for times which are about $5\%$ of the time needed to
form the gel. This phenomenology is quite close to the recent
theoretical predictions done for the violation of the fluctuation
dissipation theorem in glassy systems.
\end{abstract}

\bigskip

Many physical systems in nature are not in thermodynamic equilibrium because
they present very slow relaxation processes. A typical example of this
phenomenon is the aging of glassy materials: when they are quenched from above
their glass transition temperature $T_g$ to a temperature $T<T_g$, any
response function of these systems depends on the aging time $t_a$ spent at
$T$. For example, the dielectric and elastic constants of polymers continue to
evolve several years after the quench \cite{Struick}. Because of these slow
relaxation processes, the glass is out of equilibrium, and usual
thermodynamics does not apply. However, as this time evolution is slow, some
concepts of the classical approach may be useful for understanding the glass
aging properties. A widely studied question, is how the temperature of these
systems can be defined. One possible answer comes from the study of the
deviation to the Fluctuation Dissipation Theorem (FDT) in an out of
equilibrium system (for a review see ref. \cite{Mezard,Cugliandolo,Peliti}).
In this letter we show that this approach is relevant for the study of a
sol-gel transition, where a strong violation of FDT is measured. Implications
of this observation go beyond the physics interest. Indeed FDT is used as a
tool to extract, from fluctuations measurements, several properties in
biological, chemical and physical systems \cite{bio1,bio2,surface}. Our
results indicate that before extending this smart technique to other systems,
one has to carefully ensure that these systems are in equilibrium.

In order to understand this new definition of temperature, we have
to recall the main consequences of FDT in a system which is in
thermodynamic equilibrium. We consider an observable $V$ of such a
system and its conjugate variables $q$ (examples of conjugate
variables are voltage-charge, pressure-volume etc.). The response
function $\chi_{Vq}(\omega)$, at frequency $\nu=\omega / 2 \pi$,
describes the variation $\delta V (\omega)$ of $V$ induced by a
perturbation $\delta q (\omega)$ of  $q$, that is $\chi_{Vq}
(\omega)=\delta V (\omega)/ \delta q(\omega)$ (examples of
response functions are the electric constants and compressibility
of a material). FDT relates the fluctuation spectral density of
$V$ to the response function $\chi_{Vq}$ and the temperature T of
the system:

\begin{equation}
S(\omega) = { 2 k_B \ T \over \pi \omega } {\it
Im}\left[\chi_{Vq}(\omega) \right] \label{FDT}
\end{equation}

where $S(\omega)=<|V(\omega)|^2>$ is the fluctuation spectral density of $V$,
$k_B$ is the Boltzmann constant, ${\it Im}\left[ \chi_{Vq}(\omega) \right]$ is
the imaginary part of $\chi_{Vq}(\omega)$. Textbook examples of FDT are
Nyquist's formula relating the voltage noise to the electrical resistance and
the Einstein's relation for Brownian motion relating the  particle diffusion
coefficient  to the fluid viscosity \cite{book}.

When the system is not in equilibrium FDT may fail. For example
violations, of about a factor of 2, of eq.\ref{FDT} have been
observed in the density fluctuations of polymers in the glassy
phase \cite{Wendorff}. The first to propose that the study of the
FDT violations were relevant for glassy systems was Sompolinsky
\cite{Sompolinsky}. This idea, which was generalized in the
context of weak turbulence \cite{Shraiman}, has been recently
reconsidered by Cugliandolo and Kurchan \cite{Kurchan} and
successively tested in many analytical and numerical models of
 glass dynamics
 \cite{Peliti},\cite{Parisi}-\cite{Berthier}.
Let us briefly recall the main and general findings of these
models. Because of the slow dependence on $t_a$ of the responses
functions, it has been proposed that eq.\ref{FDT} can be used to
define an effective temperature of the system, specifically:

\begin{equation}
T_{eff} (t_a, \omega) = { S(t_a, \omega) \ \pi \omega \over  {\it
Im}\left[ \chi_{Vq}(t_a, \omega) \right] \ 2 k_B }
 \label{Teff}
\end{equation}

It is clear that if eq.\ref{FDT} is satisfied $T_{ef\!f}=T$, otherwise
$T_{ef\!f}$ turns out to be a decreasing function of $t_a$ and $\omega$. The
physical meaning of eq.\ref{Teff} is that there is a time scale (for example
$t_a$) which allows to separate the fast processes from the slow ones. In
other words the low frequency modes relax towards the equilibrium value much
slower than the high frequency ones which rapidly relax to the temperature of
the thermal bath. Therefore it is conceivable that the slow frequency modes
keep memory of higher temperatures for a long time and for this reason their
temperature should be higher than that of the high frequency ones. This
striking behavior has been observed in several numerical  models of aging
\cite{Peliti},\cite{Parisi}-\cite{Berthier}. Further analytical and numerical
studies of simple models show that eq.\ref{Teff} is a good definition of
temperature in the thermodynamic sense \cite{Cugliandolo,Peliti}. In spite of
the large amount of theoretical studies there are only a few experiments which
show a violation of FDT in real materials \cite{Wendorff,Grigera}. However
these measurements are done at a single frequency and there is no idea on how
the temperature relaxes as function of time and frequency. The experimental
analysis of the dependence of $T_{eff}(\omega,t_a)$ on $\omega$ and $t_a$ is
very useful to distinguish among different models of aging: FDT violations are
model dependent \cite{Peliti},\cite{Parisi}-\cite{Berthier}.

For these reasons we have experimentally studied the violation of
eq.\ref{FDT} during a sol-gel transition in Laponite RD
\cite{Laponite}, a synthetic clay consisting of discoid charged
particles. It disperses rapidly in water to give gels even for
very low mass fraction. Physical properties of this preparation
evolves for a long time, even after the sol-gel transition, and
have shown many similarities with standard glass aging
\cite{Kroon,Bonn}. Recent experiments \cite{Bonn} have even proved
that the structure function of Laponite at low concentration (less
than $3 \%$ mass fraction) is close to that of a colloidal glass.
For this reason we call our gel a colloidal glass.

In our experiment we measure the time evolution of the Laponite electrical
properties during the sol-gel transition. The solution is used as a conductive
liquid between the two golden coated electrodes of a cell (see fig.1). The
Laponite solution is prepared in a clean $\mathrm{N_2}$ atmosphere to avoid
$\mathrm{CO_2}$ and $\mathrm{O_2}$ contamination, which perturbs the
electrical measurements. Laponite particles are dissolved at a concentration
of $2.5 \%$ mass fraction in pure water under vigorous stirring during $20
min$. To avoid the existence of any initial structure in the sol, we pass the
solution through a $1 \mu m$ filter when filling our cell. This instant define
the origin of the aging time $t_a$ (the filling of the cell takes roughly two
minutes, which can be considered the maximum inaccuracy of $t_a$). The sample
is then sealed so that no pollution or evaporation of the solvent can occur.
At this concentration, the gelation time is about $500 h$ \cite{Kroon}, and we
only study the beginning of this sol-gel transition.

\begin{figure}

  \centerline{\epsfxsize=0.6\linewidth \epsffile{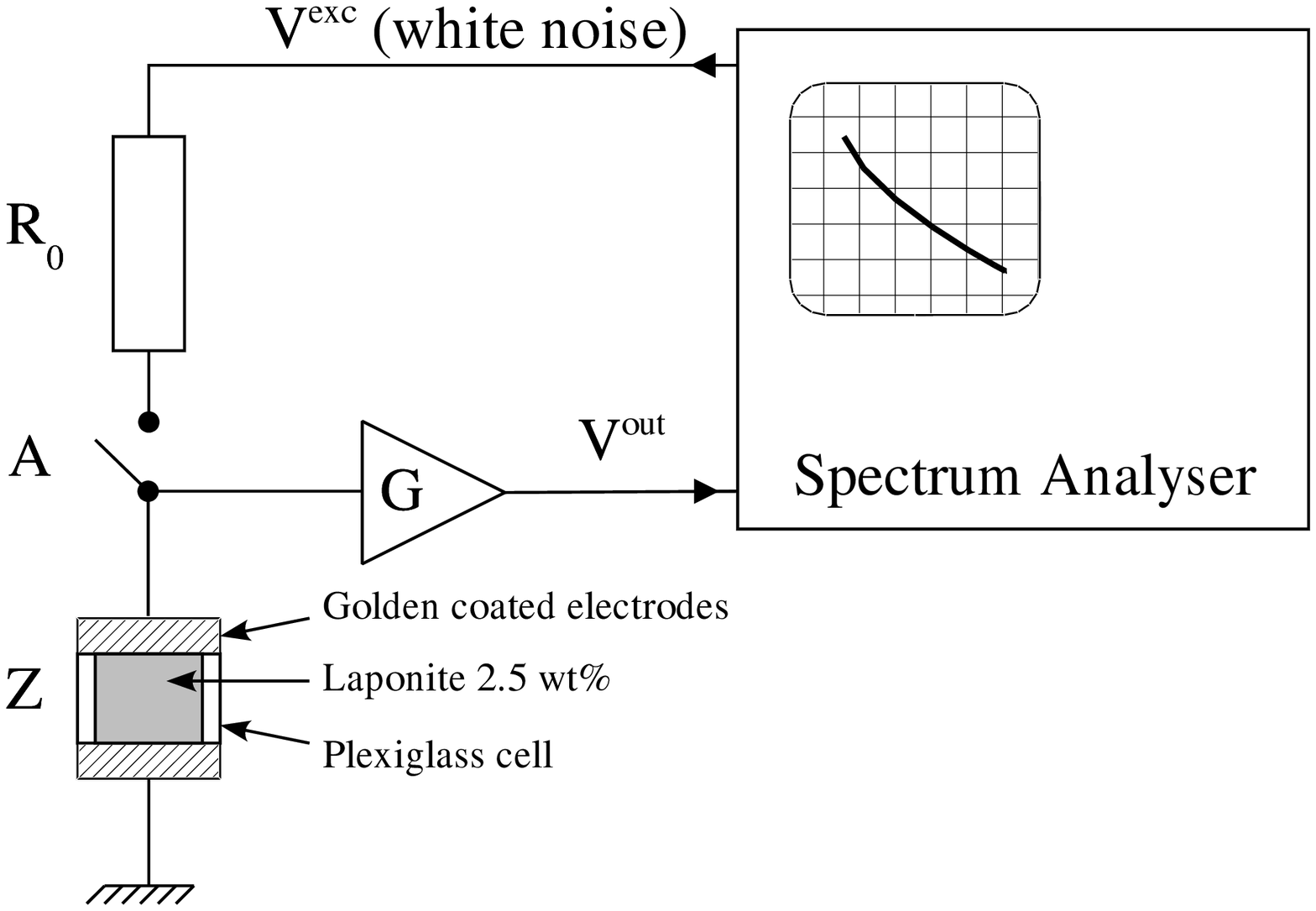}}
  \caption{{\bf Experimental set-up} The impedance under test $Z$ is a
cell (diameter $7cm$, thickness $3cm$) filled with a $2.5 wt\%$
Laponite sol. The electrodes of the cell are golden coated to
avoid oxidation. One of the two electrodes is grounded whereas the
other is connected to the entrance of a low noise  voltage
amplifier characterized by a voltage amplification $G$. With a
spectrum analyzer, we alternately record the frequency response
$F\!R(\omega)=<V^{out}/V^{exc}>$ (switch $A$ closed) and the
spectrum $S(\omega)=<|{V^{out}}|^2>$ (switch $A$ opened). The
input voltage $V^{exc}$ is a white noise excitation, thus from
$F\!R(\omega)$ we derive the impedance $Z(\omega)$ as a function
of $\omega$, that is $Z(\omega)=R_0 / (G / F\!R(\omega) - 1)$;
whereas from $S(\omega)$, we can estimate the voltage noise of
$Z$, specifically $S_Z(\omega) = [S(\omega)-S_a(\omega)]/G^2$
where $S_a(\omega)$ is the noise spectral density of the
amplifier} \label{measurement}
\end{figure}

The two electrodes of the cell are connected  to  our measurement
system, where we alternately record the cell electrical impedance
$Z(t_a,\omega)$ and the  voltage noise density $S_{Z}(t_a,\omega)$
(see Fig.\ref{measurement}). Taking into account that  in this
configuration ${\it Im }\left[ \chi_{Vq}(t_a, \omega)
\right]=\omega {\it Re}\left[ Z(t_a, \omega) \right]$, one obtains
from eq.\ref{Teff} that the effective temperature of the Laponite
solution as a function of the aging time and frequency is:

\begin{equation}
T_{eff}(t_a,\omega)=\pi S_{Z}(t_a,\omega) / 2 k_B {\it Re}\left[
Z(t_a,\omega)\right]
 \label{TLap}
\end{equation}

which is an extension of the Nyquist formula.

\begin{figure}
  {\bf (a)} \\
  \centerline{\epsfysize=0.5\linewidth \epsffile{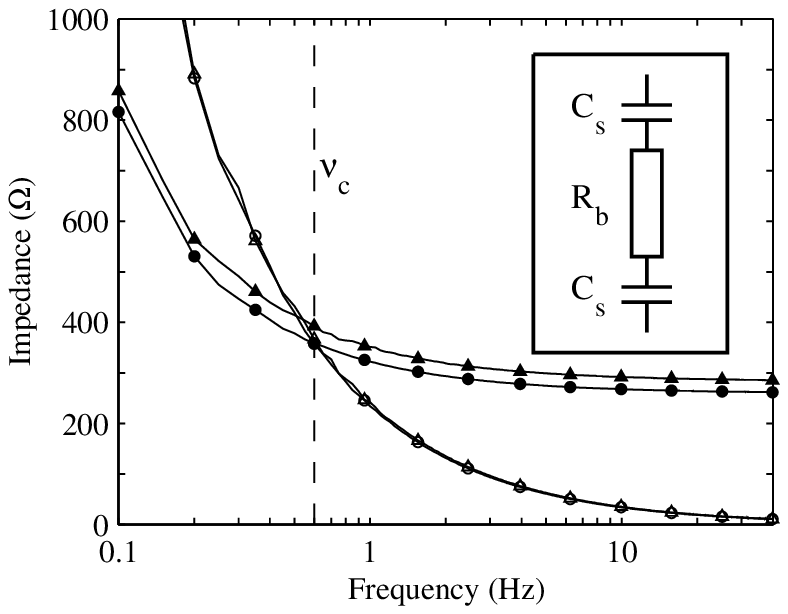}}
  {\bf (b)} \\
  \centerline{\epsfysize=0.5\linewidth \epsffile{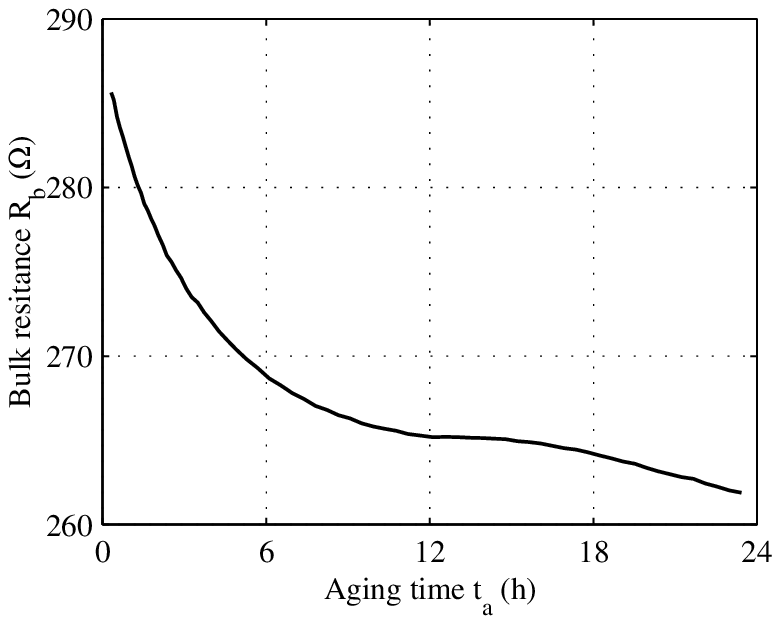}}
  \caption{{\bf The response function} (a) Frequency dependence of a sample impedance for 2
different aging time: $t_a=0.3h$, real ($\blacktriangle$) and
imaginary ($\vartriangle$) part; $t_a=24h$, real ($\bullet$) and
imaginary ($\circ$) part. The equivalent circuit for the cell
impedance is shown in the inset: $Z$ is the sum of a resistive
volume $R_b$ and a capacitive interface $C_s$ between the Laponite
solution and the electrodes.  The increase of $Re(Z)$ toward small
frequencies ($\nu<\nu_c$) is due to the dissipative part of the
capacitance (loss tangent $\simeq 0.2$). For $\nu>\nu_c$ the
impedance of the cell is dominated by the bulk resistance $R_b$.
(b) Time evolution of the bulk resistance. This long time
evolution is the signature of the aging of the sol. In spite of
the decreasing mobility of Laponite particles in solution during
the gelation, the electrical conductivity increases. This behavior
is consistent with a decreasing of the effective temperature of
the impedance.}\label{response}
\end{figure}

In Fig.\ref{response}(a), we plot the real and imaginary part of
the impedance as a function of the frequency $\nu$, for a typical
experiment. The response of the sample is the sum of 2 effects:
the bulk is purely conductive, the ions of the solution follow the
forcing field, whereas the interfaces between the solution and the
electrodes give mainly a capacitive effect due to the presence of
the Debye layer\cite{hunter}. This behavior has been validated
using a four-electrode potentiostatic technique \cite{electrochem}
to make sure that the capacitive effect is only due to the
surface. In order to test only bulk properties, the geometry of
the cell is tuned to push the surface contribution to low
frequencies: the cutoff frequency of the equivalent R-C circuit is
less than $0.6 Hz$. The time evolution of the resistance of one of
our sample is plotted in Fig.\ref{response}(b): it is still
decaying in a non trivial way after $24 h$, showing that the
sample has not reached any equilibrium yet. This aging is
consistent with that observed in light scattering experiments
\cite{Kroon}.

\begin{figure}
  \centerline{\epsfysize=0.5\linewidth \epsffile{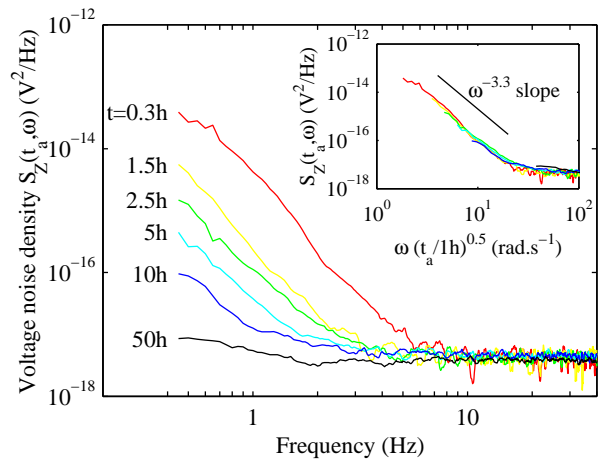}}
  \caption{{\bf Fluctuations} Voltage noise density of one sample for different
aging times. The strong increase of $S_Z$ for low frequencies is
quite well fitted by a power law $\omega^\alpha$, with $\alpha =
3.3 \pm 0.4$. This effect is a decreasing function of time, and a
good rescaling of the data with a $\omega t_a^\beta$ law can be
achieved as shown in the inset for $\beta = 0.5 \pm 0.1$.
}\label{fluctuation}
\end{figure}

\begin{figure}
{\bf (a)} \\
  \centerline{\epsfysize=0.5\linewidth \epsffile{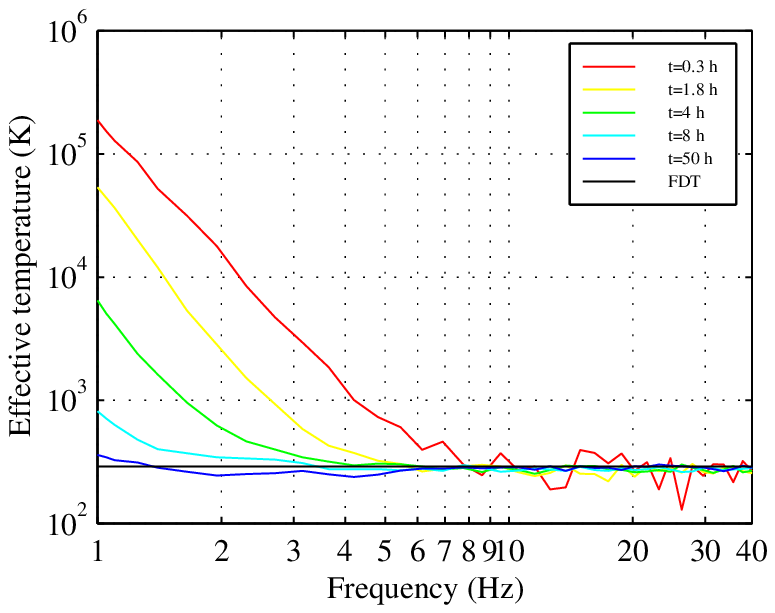}}
{\bf (b)} \\
  \centerline{\epsfysize=0.5\linewidth \epsffile{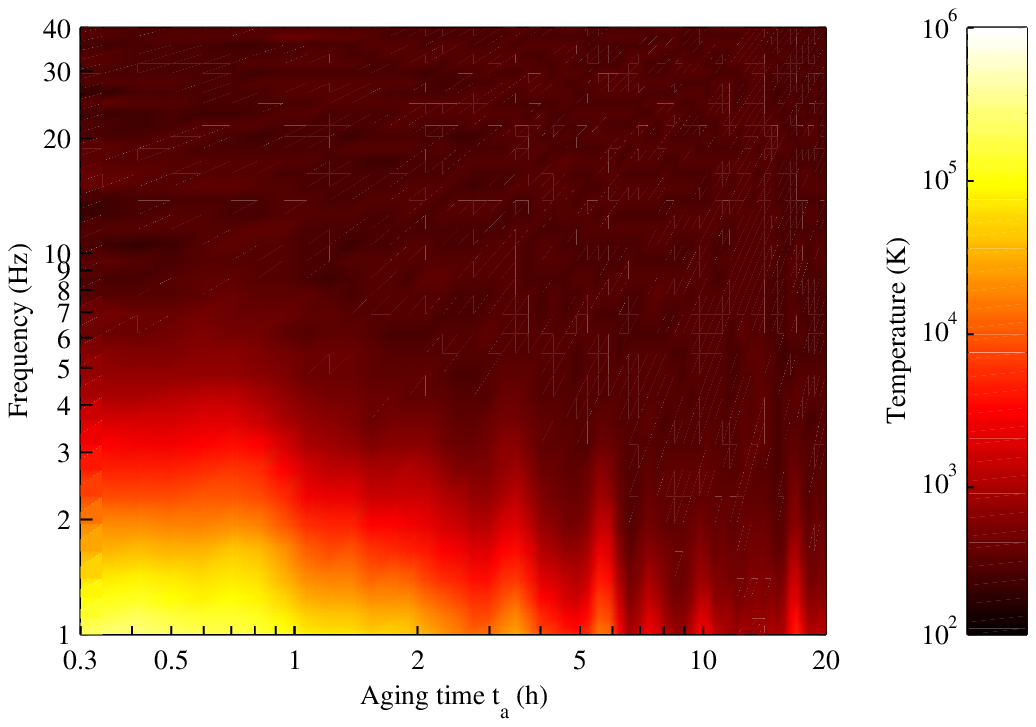}}
  \caption{{\bf Effective temperature} (a) Effective temperature
  as a function of frequency for
different aging times \cite{remarque1}. We restrict the frequency
range to $1 Hz$ to limit the surface contribution to the
measurement. As $S_Z$ in Fig.\ref{fluctuation}, $T_{ef\!f}$ is
strongly increasing for low frequencies and short aging time. (b)
Time frequency representation of the effective temperature of the
same experiment. $T_{ef\!f}$, coded with a logarithmic color
scale, is reaching huge values in the short aging time / low
frequency corner. This large violation is observed in numerical
simulations of systems presenting domain growth process which are
probably good models to describe the formation of a Laponite
gel.}\label{temperature}
\end{figure}

As the dissipative part of the impedance $Re(Z)$ is weakly time
and frequency dependent, one would expect from the Nyquist formula
that so does the voltage noise density $S_{Z}$. But as shown in
Fig.\ref{fluctuation}, FDT must be strongly violated for the
lowest frequencies and earliest times of our experiment: $S_{Z}$
changes by several orders of magnitude between highest values and
the high frequency tail \cite{1/f}. This violation is clearly
illustrated by the behavior of the effective temperature in
Fig.\ref{temperature}. For long times and high frequencies, the
FDT holds and the measured temperature is the room one ($300K$);
whereas for early times $T_{eff}$ climbs up to $3.10^5K$ at $1Hz$.
Notice that the violation extends to frequencies much larger than
the $0.6Hz$ cutoff separating the volume from the surface effects,
leaving no doubt that it takes place in the bulk. Moreover, the
scaling presented in inset of Fig.\ref{fluctuation} seems to
indicate that $T_{eff}$ can be even larger for lower frequencies
and lower aging times. Indeed, we found in all the tested samples
no evidence of a saturation of this effective temperature in our
measurement range. The observed very large value of $T_{eff}$ is
of course very striking. However the existence of infinite
$T_{eff}$ was predicted \cite{Peliti} and numerically verified
\cite{Barrat} in systems presenting domain growth process. This is
probably the way in which the Laponite solution makes the
transition towards the gel (colloidal glass) state.

In conclusion we have observed for the first time that at the beginning of the
transition from a fluid like sol to a solid like gel FDT is strongly violated.
As predicted by the theory  the amplitude and the persistence time of this
violation are decreasing functions of frequency. This large violations are
observed in numerical simulations of systems presenting domain growth process
which are probably good models to describe the formation of a Laponite gel. At
the moment we are unable to claim whether in our system $T_{eff}$ can be seen
as a temperature in the thermodynamic sense: other kind of measurements will
be necessary to do such a statement. However the observed violation is very
important from the application point of view because FDT is often invoked to
estimate the response from a measurement of fluctuations. Our results show
that this technique, although very powerful, has to be used  with a great
precaution : one has to carefully insure that the system under study has
reached equilibrium.

{\bf Acknowledgements} We thank J. Kurchan, L. Berthier and M. Carl\`a for
useful discussions and suggestions. We acknowledge L. Renaudin for technical
support. This work has been partially supported by the Programme Th\'ematique
of Region Rh\^{o}ne-Alpes.

\newpage

\newpage

\newpage

\newpage

\begin{thebibliography}{}
\bibitem{1/f} This low frequency noise cannot be confused with the
standard $1/f$ noise observed in many electronic devices. We
recall that the $1/f$ appears only when an external  current
produced by an external potential  goes through the device. In our
cell no external potential is applied.
\bibitem{remarque1} The usual representation of the effective temperature
in simulations is the slope of the response versus correlation
plot, but it is not suited for our experimental data: the system
being almost only dissipative, the response function is close to a
delta distribution, thus FDT is only one point in this
representation...
\end{thebibliography}
\end{document}